\documentclass[twocolumn,preprintnumbers,amssymb,amsmath,aps,floatfix,prl,nofootinbib,superscriptaddress]{revtex4}

\usepackage{epsfig}
\usepackage{bm}
\usepackage{amssymb}
\usepackage{amsmath}
\usepackage{color}
\usepackage{subfigure}
\usepackage{hyperref}

\begin{document}

\title{Asymmetric longitudinal flow decorrelations in proton-nucleus collisions}

\author{Xiang-Yu Wu}
\affiliation{Institute of Particle Physics and Key Laboratory of Quark and Lepton Physics (MOE), Central China Normal University, Wuhan, 430079, China}

\author{Guang-You Qin}
\email{guangyou.qin@mail.ccnu.edu.cn}
\affiliation{Institute of Particle Physics and Key Laboratory of Quark and Lepton Physics (MOE), Central China Normal University, Wuhan, 430079, China}

\begin{abstract}

We perform the first study on asymmetric longitudinal decorrelations of elliptic, triangular and quadrangular flows in proton-nucleus collisions at the LHC and RHIC energies.
To measure the longitudinal flow decorrelations for asymmetric collision systems, we propose a new set of rapidity-asymmetric flow decorrelation functions.
Our event-by-event hydrodynamic calculations show that the flow decorrelations in proton-going direction are larger than those in nucleus-going direction.
We also find that proton-nucleus collisions at RHIC have larger longitudinal flow decorrelation effects than those at the LHC.
Our study opens a new window to probe the longitudinal properties and the origin of flows in relativistic nuclear collisions.

\end{abstract}
\maketitle

{\it Introduction}. High-energy heavy-ion collisions at the Large Hadron Collider (LHC) and the Relativistic Heavy-Ion Collider (RHIC) have created strongly-interacting quark-gluon plasma (QGP).
One of the most important signatures of QGP is the strong collective flow~\cite{Ollitrault:1992bk, Adler:2003kt, Adams:2003am, Adams:2003zg, Aamodt:2010pa, ATLAS:2011ah, Chatrchyan:2012ta}, which is anisotropic in the plane transverse to the beam direction and can be quantified by the Fourier decomposition of the azimuthal angle distribution of the final-state hadrons, $dN/d\psi \propto 1 + \sum_n 2 v_n \cos[n(\psi - \Psi_n)]$, where the magnitude $v_n$ and the orientation $\Psi_n$ can be combined to define the flow vector $\mathbf{V}_n = v_n e^{in\Psi_n}$.
Relativistic hydrodynamics has been extremely successful in simulating the bulk evolution of QGP and explaining the flow data in nucleus-nucleus (AA) collisions at RHIC and the LHC~\cite{Heinz:2013th, Gale:2013da, Huovinen:2013wma, Rischke:1995ir, Romatschke:2017ejr}.
Due to the strong interaction among QGP constituents, the initial geometric anisotropies are converted to final momentum anisotropies~\cite{Gyulassy:1996br, Aguiar:2001ac, Broniowski:2007ft, Andrade:2008xh, Hirano:2009ah, Alver:2010gr, Petersen:2010cw, Qin:2010pf, Staig:2010pn, Teaney:2010vd, Schenke:2010rr, Ma:2010dv, Qiu:2011iv}.
Extensive studies of anisotropic flows in relativistic heavy-ion collisions have provided copious information on the initial states and transport properties of QGP.

Recent experiments have observed strong anisotropic flows in small systems such as proton-nucleus (pA) collisions \cite{Abelev:2012ola, Aad:2012gla, Chatrchyan:2013nka, Dusling:2015gta, Nagle:2018nvi}.
The collectivity in pA collisions can be naturally explained by hydrodynamics with fluctuating initial conditions~\cite{Bozek:2013uha, Bzdak:2013zma, Qin:2013bha, Werner:2013ipa, Bozek:2013ska, Nagle:2013lja, Schenke:2014zha, Weller:2017tsr, Zhao:2019ehg}, which indicates that the mini quark-gluon plasma might be formed in small collision systems.
However, no jet quenching has been observed in pA collisions so far~\cite{ALICE:2012mj, CMS:2015ved, ALICE:2018vuu, Albacete:2013ei, Eskola:2016oht}.
Additionally, the observed collectivity for heavy mesons in pA collisions challenges the final-state effect explanation \cite{ALICE:2017smo, CMS:2018duw, CMS:2018loe, Du:2018wsj}.
On the other hand, Color Glass condensate (CGC) framework has shown that the correlations between the incoming partons via interacting with the dense gluons in the nucleus can produce significant amount of collectivity \cite{Dusling:2012iga, Dusling:2012wy, Kovchegov:2012nd, Dusling:2013oia, Lappi:2015vta, Schenke:2015aqa, Schenke:2016lrs, Dusling:2017dqg, Dusling:2017aot, Mace:2018vwq, Mace:2018yvl, Davy:2018hsl}.
CGC can also explain heavy meson $v_2$ in pA collisions \cite{Zhang:2019dth, Zhang:2020ayy}.
It is currently still under debate whether the collectivity in pA collisions is mainly driven by initial-state gluon saturation dynamics or the final-state hydrodynamic evolution.
We also note other explanations of the collectivity in small systems \cite{Bzdak:2014dia, Lin:2015ucn, Li:2018leh, Kurkela:2018qeb}.

While tremendous efforts have been devoted to harmonic flows, many of them focused on the fluctuations in the transverse plane and usually assumed the boost-invariance approximation for the initial conditions and the evolution dynamics of QGP.
In fact, the fireball produced in relativistic nuclear collisions is not longitudinally boot-invariant, and even in a single event, $\mathbf{V}_n$ fluctuates along the rapidity direction, i.e., $\mathbf{V}_n(\eta_1) \neq  \mathbf{V}_n(\eta_2)$ for $\eta_1 \neq \eta_2$.
Such longitudinal fluctuation and decorrelation effects have been observed by experiments \cite{CMS:2015xmx, ALICE:2016tlx,CMS:2017xnj,  ATLAS:2017rij, Nie:2019bgd, ATLAS:2020sgl}.
Various studies have shown that the longitudinal fluctuations and decorrelations of anisotropic flows are mainly driven by the fluctuations of initial geometry along the rapidity direction \cite{Petersen:2011fp, Xiao:2012uw, Pang:2012uw, Rybczynski:2013yba, Pang:2014pxa, Jia:2014vja, Jia:2014ysa, Bozek:2015bha, Bozek:2015bna, Jia:2015jga, Denicol:2015nhu,  Broniowski:2015oif, Bozek:2015swa, Bozek:2015tca, Pang:2015zrq, Ke:2016jrd, Schenke:2016ksl,  Jia:2017kdq, Shen:2017bsr, Bozek:2017qir, Wu:2018cpc, Bozek:2018nne, Pang:2018zzo, Shen:2020jwv, Xu:2020koy, Behera:2020mol, Sakai:2020pjw, Schlichting:2020wrv, Cimerman:2021gwf, He:2020xps, Jia:2020tvb}.
To further our understanding of longitudinal dynamics of QGP, it is crucial to investigate in detail the longitudinal fluctuations and decorrelations in different collision systems and across different colliding energies.
Along this direction, ATLAS Collaboration has recently measured the longitudinal flow decorrelations in Xe-Xe collisions at the LHC, providing an important lever-arm to study the longitudinal structure of the fireball produced in heavy-ion collisions \cite{ATLAS:2020sgl}.

This work investigates a new phenomenon: rapidity-asymmetric flow decorrelations in pA collisions.
Proton-nucleus collisions provide a unique environment for studying the dynamics of relativistic nuclear collisions.
First, the origin of the collectivity of heavy and light particles in pA collisions is still unsettled.
Another particularly compelling feature of pA collisions is that the system is asymmetric in forward and backward rapidity directions.
To measure the longitudinal flow decorrelations for such asymmetric systems, we propose a new set of rapidity-asymmetric flow decorrelation functions.
By performing (3+1)-dimensional event-by-event hydrodynamic simulations, we compute the asymmetric longitudinal decorrelations for elliptic, triangular and quadrangular flow ($v_{2,3,4}$) in proton-nucleus collisions at the LHC and RHIC energies.
Our result shows that the flow decorrelations in proton-going direction are larger than those in nucleus-going direction.
We also find that pA collisions at RHIC have larger longitudinal flow decorrelation effects than at the LHC.
Our work opens a new window to probe the longitudinal properties and the origin of flows in proton-nucleus collisions.

{\it Relativistic hydrodynamic simulations}. To study the longitudinal fluctuations and flow decorrelation in relativistic nuclear collisions, we use the (3+1)-dimensional relativistic hydrodynamics model CLVisc \cite{Pang:2012he, Pang:2018zzo, Wu:2021fjf} to simulate the dynamical evolution of the fireball produced in the collisions.
The initial conditions for hydrodynamics are provided by the AMPT model with the string-melting mechanism \cite{Lin:2004en, Zhang:2005ni}.
Using the partons from the AMPT model, the energy-momentum tensor $T^{\mu\nu}$ at the initial proper time $\tau_0$ is constructed as follows:
\begin{align}
& T^{\mu\nu} (\tau_0,x,y,\eta_s) = K \sum_i \frac{p^{\mu}_ip^{\nu}_i}{p^{\tau}_i}\frac{1}{\tau_0\sqrt{2\pi \sigma^2_{\eta_s}}}\frac{1}{2\pi \sigma^2_{r}} \nonumber \\
& \times  \exp\left[-\frac{(x-x_i)^2+(y-y_i)^2}{2\sigma_r^2}-\frac{(\eta_s-\eta_{si})^2}{2\sigma^2_{\eta_s}}\right] \,.
\end{align}
Here $p^{\mu} = [m_{T}{\rm cosh}(Y-\eta_{s}),p_{x},p_{y},\frac{1}{\tau_0} m_{T}{\rm sinh}(Y-\eta_{s})]$ is the parton's four-momentum,
with $Y$, $\eta_{s}$, $m_{T}$ being its rapidity, space-time rapidity and transverse mass, respectively.
The parameters for Gaussian smearing functions are taken as $\sigma_r = 0.4$~fm and $\sigma_{\eta_s} = 1.4$.
The initial proper time is taken to be $\tau_0=0.2$~fm/c for p-Pb collisions at $\sqrt{s_{\rm NN}} = 5.02$~TeV and $\tau_0 = 0.4$~fm/c for p-Au collisions at $\sqrt{s_{\rm NN}} = 200$~GeV.
The normalization parameter $K$ is tuned by comparing our hydrodynamics results to the charged hadron $dN_{\rm ch}/d\eta$ data \cite{ATLAS:2015hkr, PHENIX:2018hho}.
In the AMPT model, the elastic parton cross section is set as $\sigma = 3$~ mb, and the Lund string fragmentation parameters are chosen to be $a=0.3$ and $b = 0.15$ \citep{Bzdak:2014dia}.

Given the initial conditions, we numerically solve the energy-momentum conservation equation and the 2nd order Israel-Stewart-like equations for shear stress tensor, using the equation of state from s95p-PCE-v0 \citep{Huovinen:2009yb}.
The specific shear viscosity is taken to be $\eta/s=0.16$ for p-Pb collisions at $\sqrt{s_{\rm NN}} = 5.02$~TeV and $\eta/s=0.08$ for p-Au collisions at $\sqrt{s_{\rm NN}} = 200$~GeV.
After hydrodynamics evolution, the momentum distributions of the produced hadrons are obtained via the Cooper-Frye formula, with the freezeout temperature taken as $T_f=137$~MeV.

In this event-by-event study, we determine the collision centrality using the initial parton multiplicity from the AMPT model by running $10^5$ minimum-bias events.
For each centrality class, we run 2000 hydrodynamics events for analyzing flow and flow decorrelation observables.

{\it Rapidity-asymmetric flow decorrelations}.  In this work, we use the $\mathbf{Q}_n$ vector notation to quantify the $n$-th order harmonic flow,
\begin{align}
\mathbf{Q}_n(\eta) = \frac{1}{N}\sum^N_{i=1} e^{in\psi_i}\,,
\end{align}
where the sum runs over all particles in a specific pseudorapidity $\eta$ bin, and $\psi_i$ is the azimuthal angle of particle momentum.
In practice, we use the smooth particle spectra from hydrodynamic simulations to compute $\mathbf{Q}_n$,
\begin{align}
\mathbf{Q}_n(\eta) = \frac{\int \text{exp}(in\psi) \frac{dN}{d\eta dp_Td\psi }dp_Td\psi}{\int \frac{dN}{d\eta dp_Td\psi }dp_Td\psi}\,.
\end{align}
Then the $\mathbf{Q}_n$ vector is the same as the flow vector $\mathbf{V}_n$.

To study the longitudinal decorrelation of harmonic flows between two different rapidity bins, CMS Collaboration proposed a reference rapidity method and define the following flow decorrelation function \cite{CMS:2015xmx},
\begin{align}
r_n(\eta|\eta_{\rm r}) = \frac{\langle \mathbf{Q}_n(\eta) \mathbf{Q}_n^*(-\eta_{\rm r})\rangle}{\langle \mathbf{Q}_n(-\eta)\mathbf{Q}_n^*(-\eta_{\rm r})\rangle} \,,
\label{cms_rn}
\end{align}
where $\langle \cdots \rangle$ is the average over many events.
Note that in the above definition, $\eta>0$ and $\eta_{\rm r} >0$ are usually assumed.
The function $r_n(\eta|\eta_{\rm r})$ measures the decorrelation effect between two symmetric rapidity bins ($\eta$ and $-\eta$) by comparing each of them to the reference rapidity bin $-\eta_{\rm r}$, which is usually chosen to be large to remove short range correlations.
It is easy to see that $r_n(\eta|\eta_{\rm r}) = r_n(-\eta|-\eta_{\rm r})$ for symmetric collision systems.
Later ATLAS \cite{Jia:2017kdq, ATLAS:2017rij} generalized the CMS definition using the $k$-th moment of $\mathbf{Q}_n$ to define, $r_{n,k}(\eta|\eta_{\rm r}) = \frac{\langle \mathbf{Q}_n^k(\eta) \mathbf{Q}_n^{*k}(-\eta_{\rm r})\rangle}{\langle \mathbf{Q}_n^k(-\eta)\mathbf{Q}_n^{*k}(-\eta_{\rm r})\rangle}$.

%%%%%%%%%%%%%%%%%%%%%%%%%%%%%%%%%%%%%%
\begin{figure*}[thb]
\includegraphics[width=0.315\linewidth]{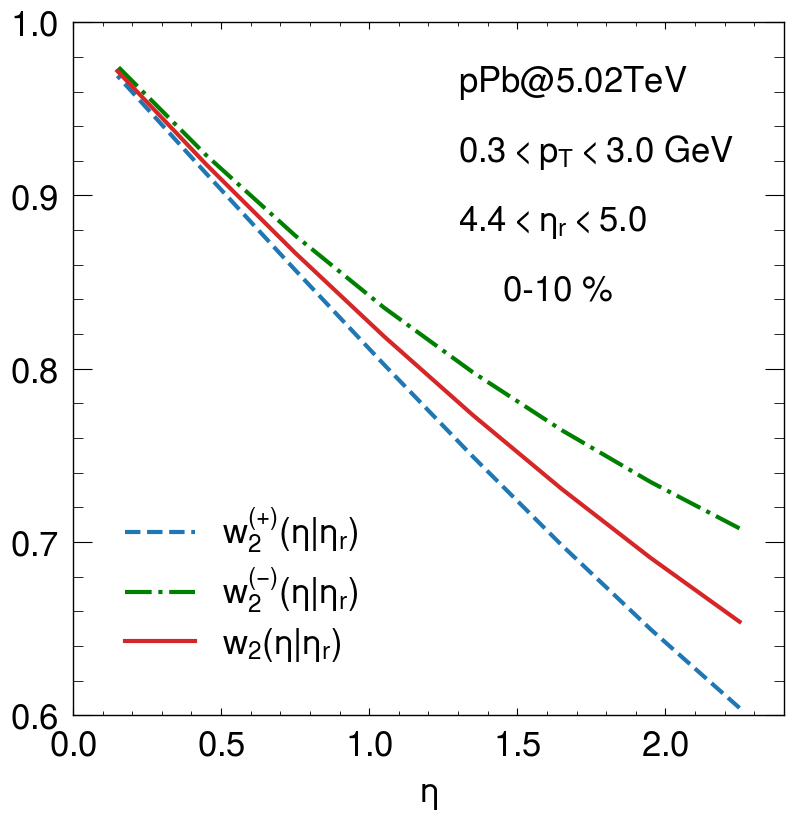}
\includegraphics[width=0.315\linewidth]{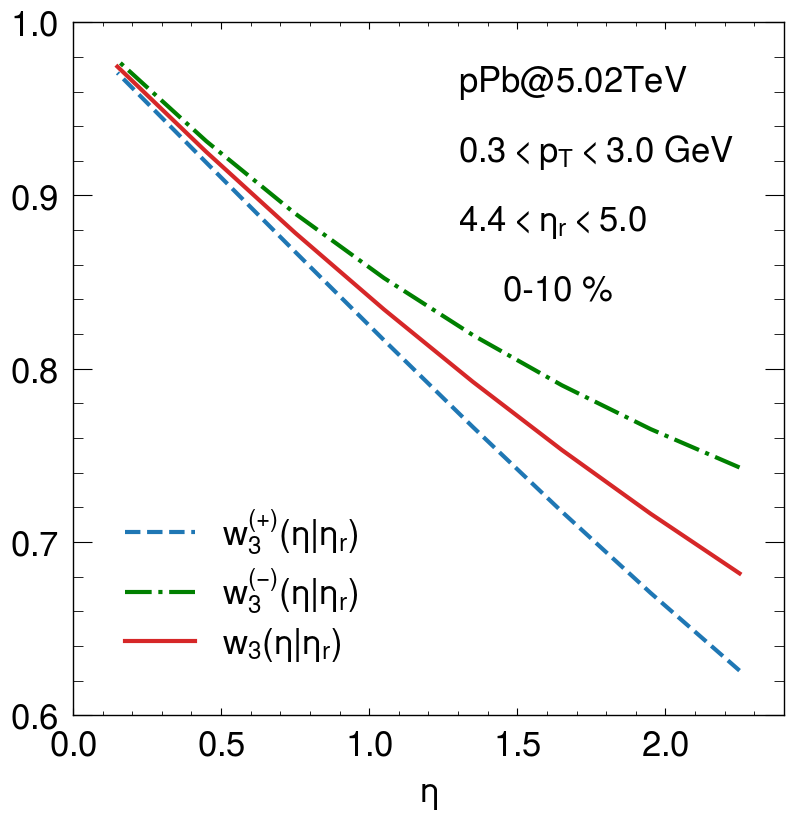}
\includegraphics[width=0.315\linewidth]{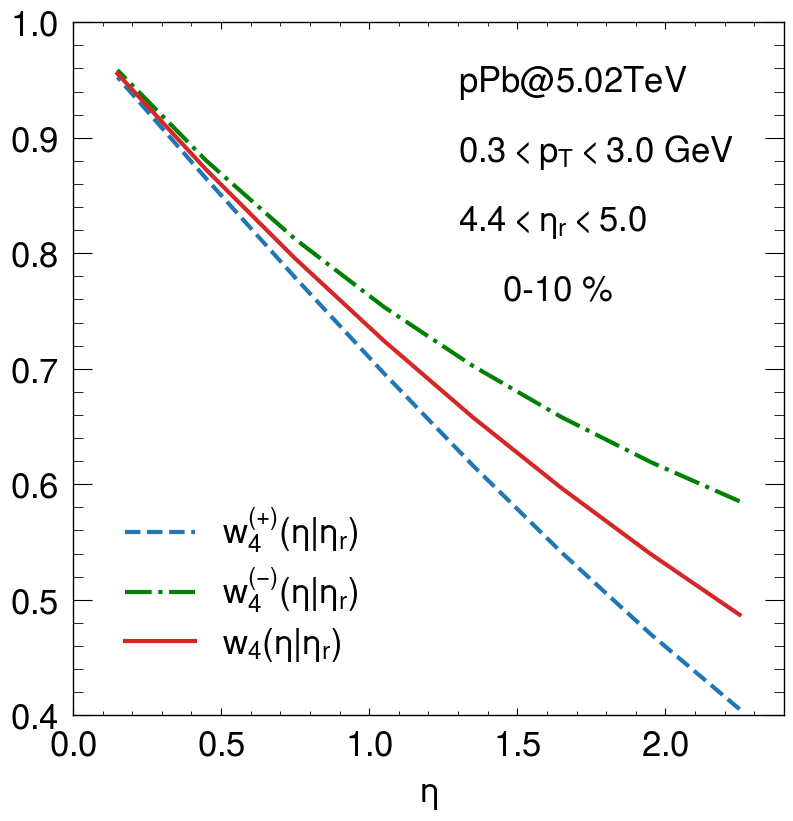}
\caption{
The longitudinal flow decorrelation functions $w_n^{(\pm)}(\eta|\eta_{\rm r})$ and $w_n(\eta|\eta_{\rm r})$ with $n=2, 3, 4$ as a function of $\eta$ obtained from event-by-event hydrodynamics simulations for $0-10\%$ p-Pb collisions at $\sqrt{s_{\rm NN}}=5.02$~TeV.
}
\label{wn_pPb_lhc}
\end{figure*}
%%%%%%%%%%%%%%%%%%%%%%%%%%%%%%%%%%%%%%

%%%%%%%%%%%%%%%%%%%%%%%%%%%%%%%%%%%%%%
\begin{figure*}[thb]
\includegraphics[width=0.315\linewidth]{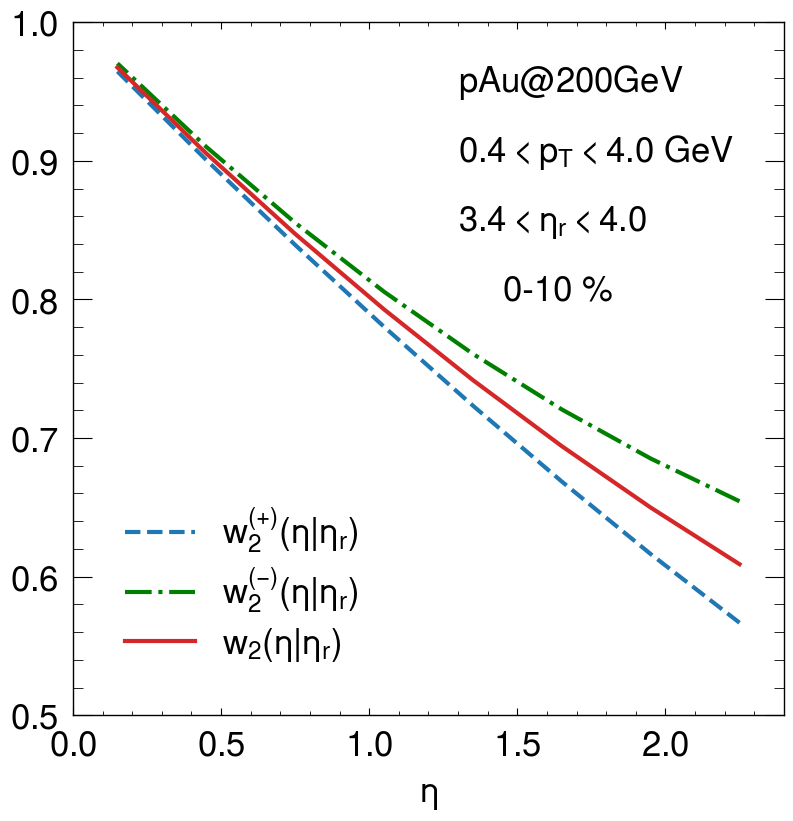}
\includegraphics[width=0.315\linewidth]{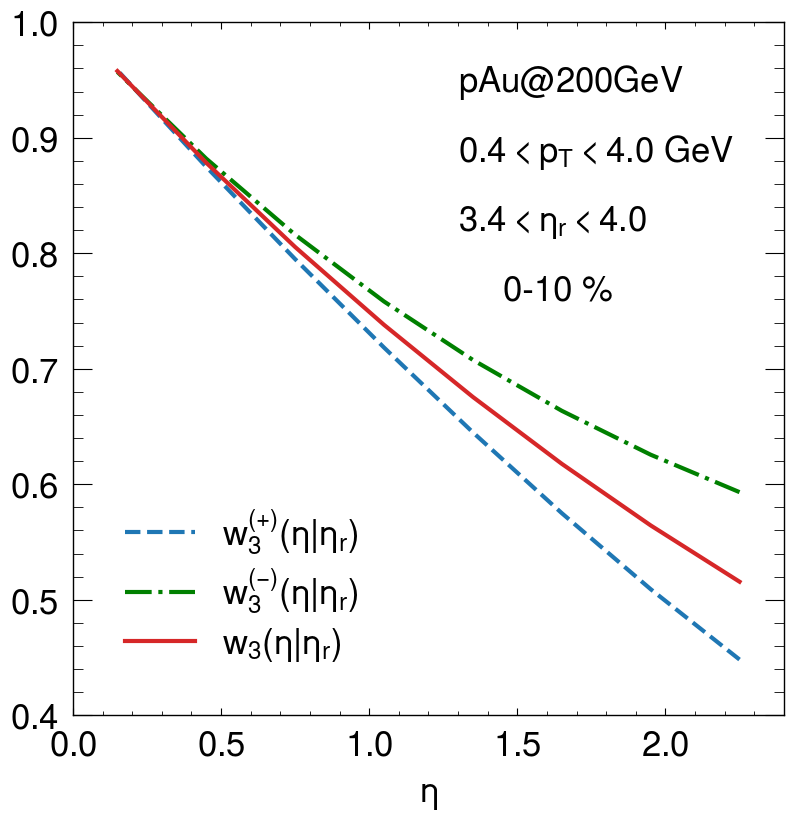}
\includegraphics[width=0.315\linewidth]{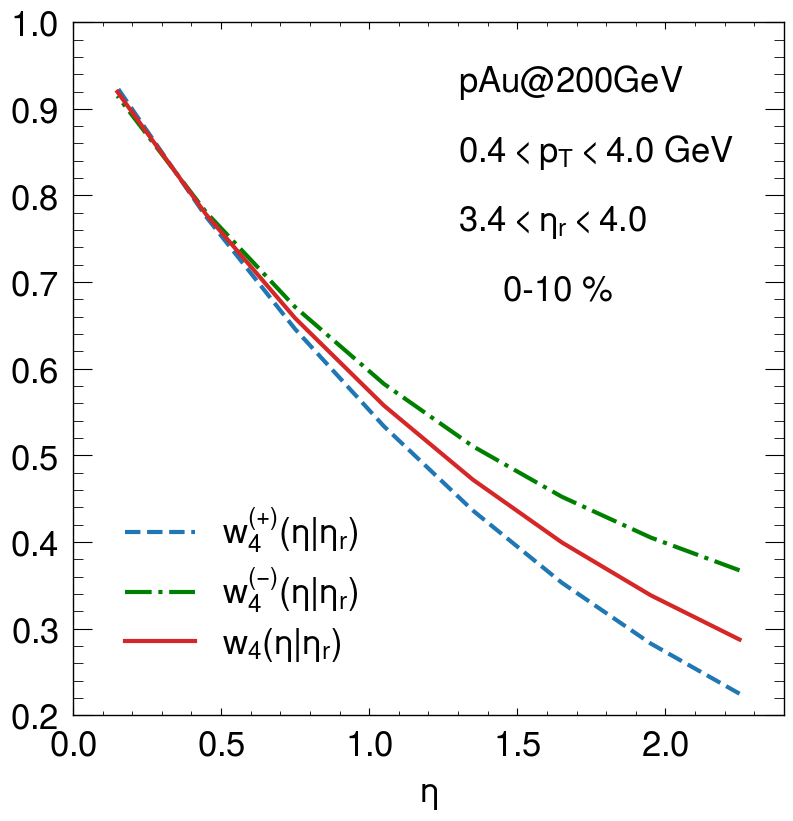}
\caption{
The longitudinal flow decorrelation functions $w_n^{(\pm)}(\eta|\eta_{\rm r})$ and $w_n(\eta|\eta_{\rm r})$ with $n=2, 3, 4$ as a function of $\eta$ obtained from event-by-event hydrodynamics simulations for $0-10\%$ p-Au collisions at $\sqrt{s_{\rm NN}}=200$~GeV.
}
\label{wn_pAu_rhic}
\end{figure*}
%%%%%%%%%%%%%%%%%%%%%%%%%%%%%%%%%%%%%%

The above definition in Eq.~(\ref{cms_rn}) works well for symmetric systems such as Pb+Pb and Au+Au collisions.
However, for asymmetric systems such as p-Pb collisions, $r_n(\eta|\eta_{\rm r}) \neq r_n(-\eta|-\eta_{\rm r})$; they cannot describe the flow decorrelations between two symmetric rapidity bins $\eta$ and $-\eta$.
To measure the longitudinal flow decorrelations in asymmetric collision systems, we propose the following rapidity-asymmetric flow decorrelation functions:
\begin{align}
{w}_n^{(+)}(\eta|\eta_{\rm r})
&= \frac{\langle \mathbf{Q}_n(\eta) \mathbf{Q}_n^*(-\eta_{\rm r})\rangle}{\langle \mathbf{Q}_n(0)\mathbf{Q}_n^*(-\eta_{\rm r})\rangle} \frac{\langle \mathbf{Q}_n(0) \mathbf{Q}_n^*(\eta_{\rm r})\rangle}{\langle \mathbf{Q}_n(\eta)\mathbf{Q}_n^*(\eta_{\rm r})\rangle} \,,
\end{align}
where $\eta > 0$ and $\eta_{\rm r} > 0$ are assumed.
The function ${w}_n^{(+)}(\eta|\eta_{\rm r})$ measures the decorrelation effect between a forward rapidity bin $\eta$ and midrapidity by comparing their correlations with the reference rapidity bins $\pm \eta_{\rm r}$.
Similarly, we define the decorrelation function between a backward rapidity bin $-\eta$ and midrapidity as follows:
\begin{align}
{w}_n^{(-)}(\eta|\eta_{\rm r})
&= \frac{\langle \mathbf{Q}_n(-\eta) \mathbf{Q}_n^*(\eta_{\rm r})\rangle}{\langle \mathbf{Q}_n(0)\mathbf{Q}_n^*(\eta_{\rm r})\rangle} \frac{\langle \mathbf{Q}_n(0) \mathbf{Q}_n^*(-\eta_{\rm r})\rangle}{\langle \mathbf{Q}_n(-\eta)\mathbf{Q}_n^*(-\eta_{\rm r})\rangle} \,.
\end{align}
Again $\eta > 0$ and $\eta_{\rm r} > 0$ are assumed.
For asymmetric systems, one typically has ${w}_n^{(+)}(\eta|\eta_{\rm r}) \neq {w}_n^{(-)}(\eta|\eta_{\rm r})$.

Given the above two types of rapidity-asymmetric flow decorrelation functions, one may perform the geometric average and define the following rapidity-symmetrized decorrelation functions:
\begin{align}
{w}_n(\eta|\eta_{\rm r})
&= \sqrt{ {w}_n^{(+)}(\eta|\eta_{\rm r}) {w}_n^{(-)}(\eta|\eta_{\rm r}) }
\end{align}
It is easy to see that ${w}_n(\eta|\eta_{\rm r}) = \sqrt{r_n(\eta|\eta_{\rm r}) r_n(-\eta|-\eta_{\rm r})}$ \cite{CMS:2015xmx}.
Note that we may define the generalized rapidity asymmetric and symmetrized flow decorrelation functions ${w}_{n,k}^{(\pm)}(\eta|\eta_{\rm r})$ and ${w}_{n,k}(\eta|\eta_{\rm r})$ by replacing $\mathbf{Q}_n$ with its $k$-th moment $\mathbf{Q}_n^k$ \cite{ ATLAS:2017rij}.
Here we focus on the $k=1$ case.

Since the longitudinal flow decorrelation functions ${w}_n^{(\pm)}$ and ${w}_n(\eta|\eta_{\rm r})$ around midrapidity are almost linear in $\eta$, one may parameterize them as follows:
\begin{align}
{w}_n^{(\pm)}(\eta|\eta_{\rm r}) = 1 - 2 g_n^{(\pm)} \eta\,,
{w}_n(\eta|\eta_{\rm r}) = 1 - 2 {g}_n \eta \,.
\end{align}
The slope parameters satisfy: ${g}_n \approx \frac{1}{2}(g_n^{(+)} + g_n^{(-)})$ around midrapidity.
In practice, we follow ATLAS \cite{ ATLAS:2017rij} and measure the slope parameters as follows:
\begin{align}
g_n^{(\pm)} = \frac{\sum_i [1 - w_n^{(\pm)}(\eta_i|\eta_{\rm r})]\eta_i}{2\sum_i \eta_i^2}\,, {g}_n = \frac{\sum_i [1 - {w}_n(\eta_i|\eta_{\rm r})]\eta_i}{2\sum_i \eta_i^2} \,,
\end{align}
where $\sum_i$ runs over all $\eta$ bins in a given rapidity range which we choose as $0<\eta<2.4$.

%%%%%%%%%%%%%%%%%%%%%%%%%%%%%%%%%%%%%%
\begin{figure*}[thb]
\includegraphics[width=0.32\linewidth]{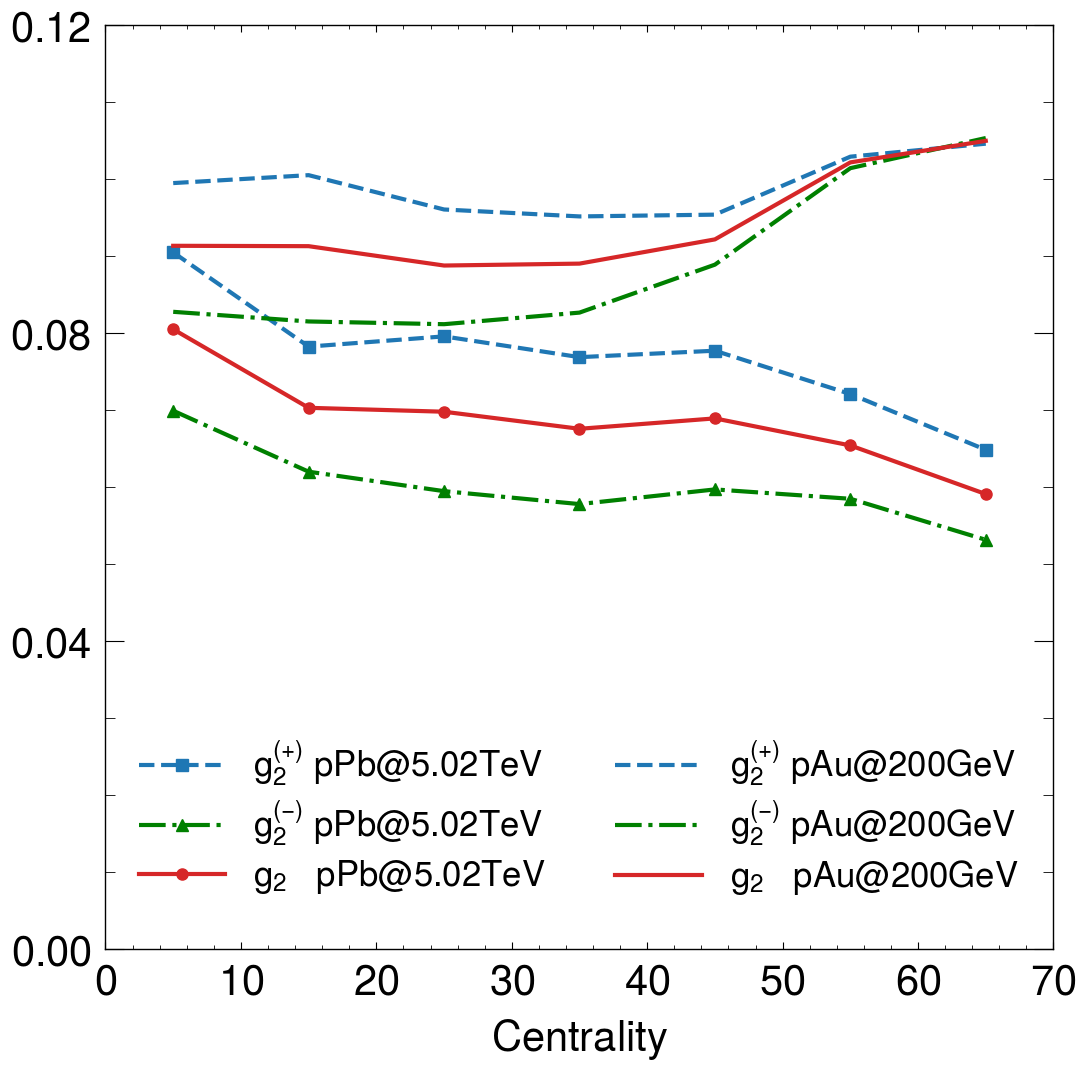}
\includegraphics[width=0.32\linewidth]{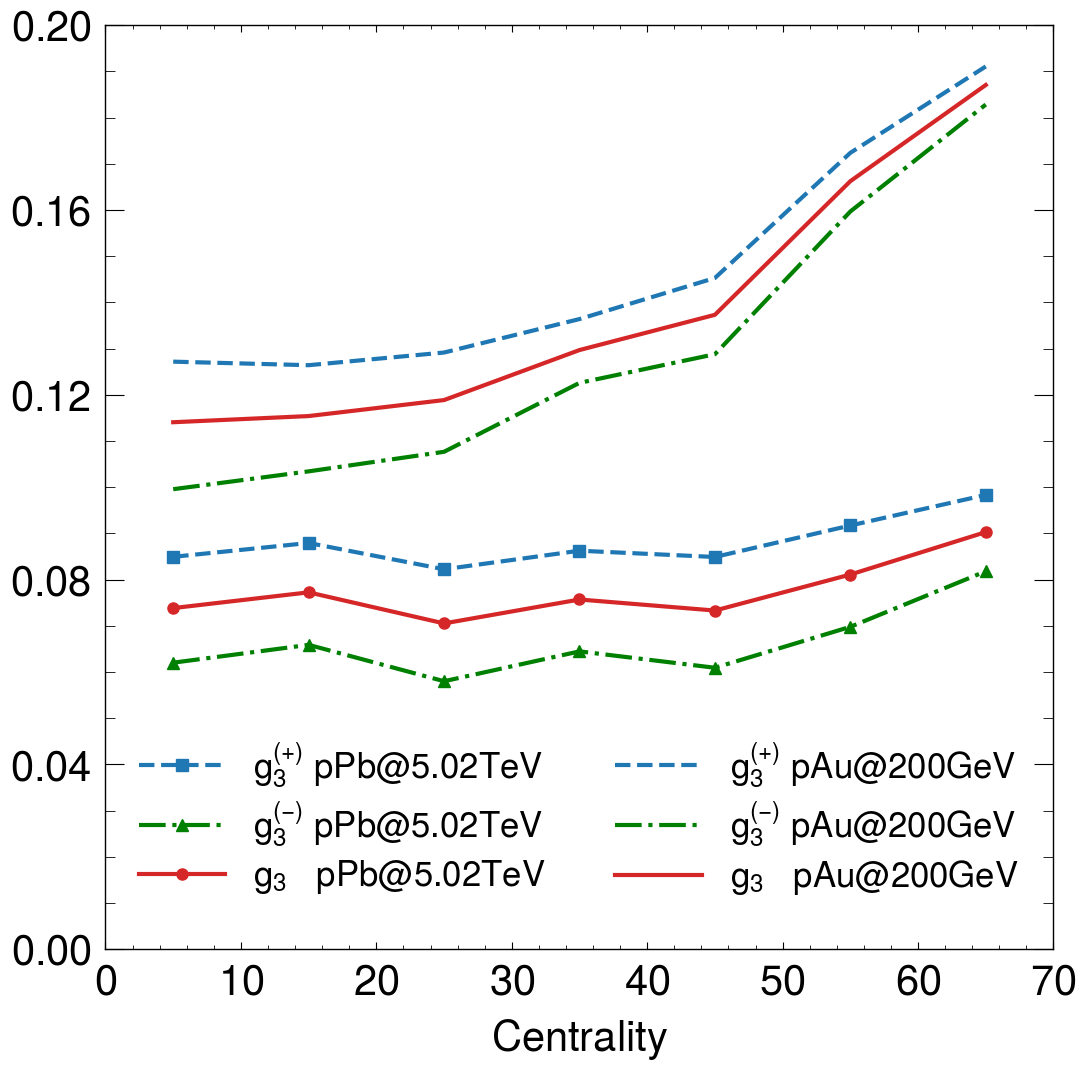}
\includegraphics[width=0.32\linewidth]{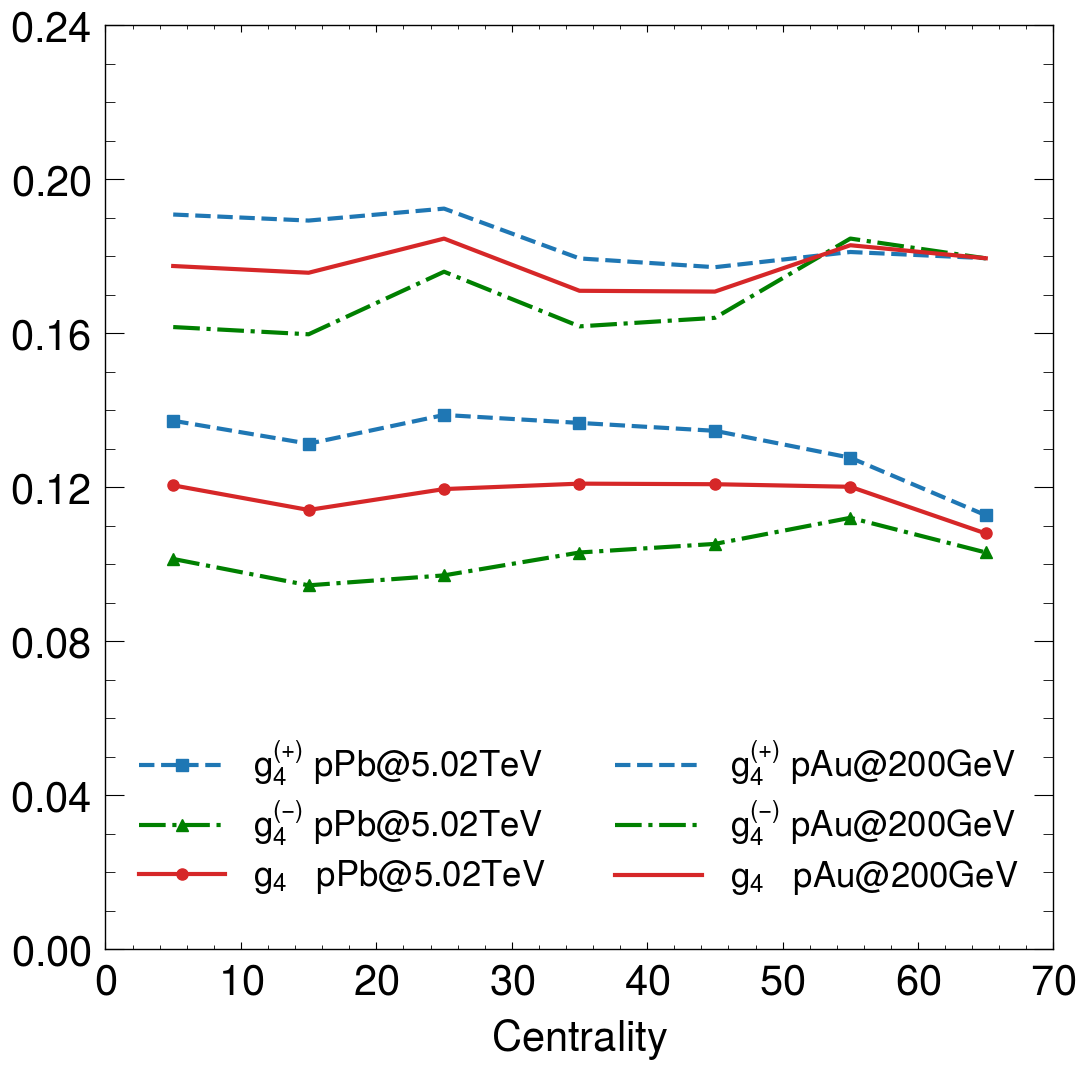}
\caption{The slope parameters $g_n^{(\pm)}$ and $g_n$ with $n=2, 3, 4$ for flow decorrelations as a function of centrality obtained from event-by-event hydrodynamics simulations for p-Pb collisions at $\sqrt{s_{\rm NN}}=5.02$~TeV and p-Au collisions at $\sqrt{s_{\rm NN}}=200$~GeV.
}
\label{gn_pA_rhic_lhc}
\end{figure*}
%%%%%%%%%%%%%%%%%%%%%%%%%%%%%%%%%%%%%%

{\it Numerical results}. Figure \ref{wn_pPb_lhc} shows the longitudinal flow decorrelation functions $w_n^{(\pm)}(\eta|\eta_{\rm r})$ and $w_n(\eta|\eta_{\rm r})$ for elliptic, triangular and quadrangular flows ($v_{2,3,4}$) as a function of pseudorapidity $\eta$ obtained from our event-by-event hydrodynamics simulations for $0-10\%$ p-Pb collisions at $\sqrt{s_{\rm NN}}=5.02$~TeV at the LHC.
In the analysis, we take particles with $0.3<p_T<3$~GeV and set the reference rapidity bins to be $4.4<\eta_{\rm r}<5.0$.
One can clearly see that $w_n^{(\pm)}(\eta|\eta_{\rm r})$ and $w_n(\eta|\eta_{\rm r})$ are decreasing as a function $\eta$, which means that the flow decorrelations are larger for larger rapidity gap, similar to the results in AA collisions \cite{Pang:2015zrq, Wu:2018cpc}.
Another very important observation is that $w_n^{(+)}(\eta|\eta_{\rm r}) < w_n^{(-)}(\eta|\eta_{\rm r})$, which states that the flow decorrelations in proton-going direction are larger than those in lead-going direction.
This is due to the fact that the fireball sizes in proton-going direction are typically smaller than those in lead-going direction.

In Fig. \ref{wn_pAu_rhic}, we show the flow decorrelation functions $w_n^{(\pm)}(\eta|\eta_{\rm r})$ and $w_n(\eta|\eta_{\rm r})$ with $n=2, 3, 4$ as a function of $\eta$ for $0-10\%$ p-Au collisions at $\sqrt{s_{\rm NN}}=200$~GeV at RHIC.
In the analysis, we take particles with $0.4<p_T<4$~GeV and set the reference rapidity bins to be $3.4<\eta_{\rm r}<4.0$.
First, we find very similar results for p-Au collisions as compared to p-Pb collisions: the flow decorrelations are larger for larger rapidity gap; the flow decorrelations in proton-going direction are larger than those in nucleus-going direction.
In addition, one can see that the average flow decorrelation effects are larger in p-Au collisions at RHIC compared to those in p-Pb collisions at the LHC.
This is due to the stronger violation of the longitudinal boost invariance in lower colliding energies, similar to the results in AA collisions \cite{Pang:2015zrq, Wu:2018cpc, Nie:2019bgd}.
Another possible reason could be the multiplicity difference between p-Au and p-Pb collisions.
Our numerical result shows that the multiplicity in 0-10\% p-Au collisions at RHIC is similar to that in 60-70\% p-Pb collisions at the LHC.

In Fig. \ref{gn_pA_rhic_lhc}, we show the slope parameters $g_n^{(\pm)}$ and $g_n$ for the flow decorrelation functions  $w_n^{(\pm)}(\eta|\eta_{\rm r})$ and $w_n(\eta|\eta_{\rm r})$ with $n=2,3,4$ as a function of centrality for p-Pb collisions at $\sqrt{s_{\rm NN}}=5.02$~TeV at the LHC and for p-Au collisions at $\sqrt{s_{\rm NN}}=200$~GeV at RHIC.
One can clearly see that $g_n^{(+)} > g_n^{(-)}$, which means that the flow decorrelations in proton-going direction are larger than those in nucleus-going direction.
Such forward-backward-asymmetry for longitudinal flow decorrelations can reach about 40\% at the LHC and 15\% at RHIC; it decreases from central to peripheral collisions because the difference between nucleus-going and proton-going directions diminish.
Additionally, the average flow decorrelation effects in p-Au collisions at RHIC are larger than those in p-Pb collisions at the LHC, due to the larger violation of the longitudinal boost invariance in lower colliding energies.
Another interesting observation is that the longitudinal decorrelation of $v_3$ tends to increase from central to peripheral collisions mainly due to the decreasing of the multiplicity; such effect is stronger at RHIC.
As for the longitudinal decorrelations of $v_2$ and $v_4$, the centrality dependences exhibit rather complex structure due to larger competing effect from collision geometry.

{\it Summary}. We have presented the first study on asymmetric longitudinal decorrelations of elliptic, triangular and quadrangular flows in proton-nucleus collisions at the LHC and RHIC energies.
To measure the longitudinal flow decorrelations for asymmetric nuclear collisions, we have constructed a new set of rapidity-asymmetric flow decorrelatoin functions.
Using our event-by-event hydrodynamics simulations, we have computed both rapidity-asymmetric and rapidity-symmetrized flow decorrelation functions for proton-nucleus collisions at the LHC and RHIC.
Our result shows that the flow decorrelations in proton-going direction are larger than those in nucleus-going direction.
It is also found that the longitudinal flow decorrelation effects are typically larger at RHIC than at the LHC.
These findings can be easily tested by experimental measurements.
Our study on asymmetric longitudinal flow decorrelations provides a novel tool to characterize the flow-like phenomena in proton-nucleus collisions and opens a new window to probe the origin of flows in small systems.
One important application of our work is to investigate the rapidity-asymmetric flow decorrelations in other asymmetric systems, such as deuteron-nucleus, $^3$He-nucleus collisions \cite{Nagle:2013lja, Bozek:2014cya, OrjuelaKoop:2015jss, Romatschke:2015gxa, Bozek:2015qpa, Mace:2018vwq, PHENIX:2015idk, PHENIX:2017djs, PHENIX:2018hho, PHENIX:2018lia}.
Also since the initial-state gluon saturation dynamics can generate significant amount of collectivity in small systems, it is extremely interesting to investigate rapidity-asymmetric flow decorrelations in proton-nucleus collisions within the CGC framework.
To conclude, systematic studies of longitudinal fluctuations and flow decorrelations in various symmetric and asymmetric systems across different colliding energies should provide unprecedented knowledge on the longitudinal as well as transverse dynamics of relativistic nuclear collisions.

{\it Acknowledgments}. This work is supported in part by Natural Science Foundation of China (NSFC) under Grants No. 11775095, No. 11890710, No. 11890711, No. 11935007.
Some of the calculations were performed in the Nuclear Science Computing Center at Central China Normal University (NSC$^3$), Wuhan, Hubei, China.

\bibliographystyle{h-physrev5} %reference style
\bibliography{refs_GYQ}   % data for reference

\end{document}